\documentclass[]{aastex631}
\usepackage{bm}
\usepackage{amsmath}
\usepackage{upgreek}
\usepackage{acronym}
\let\tablenum\relax	
\usepackage{siunitx}
\usepackage{url}

\newcommand{\width}{0.5\textwidth}

\newcommand{\ourrealnc}{893}
\newcommand{\ourrealc}{312}
\newcommand{\Fengrealnc}{1086}
\newcommand{\Fengrealc}{418}
\newcommand{\ourminnc}{279}
\newcommand{\ourminc}{80}
\newcommand{\ourmaxnc}{2225}
\newcommand{\ourmaxc}{779}
\newcommand{\offaxisrationc}{88}
\newcommand{\offaxisratioc}{98}

\acrodef{SVD}{singular value decomposition}

\DeclareSIUnit{\erg}{erg}
\DeclareSIUnit{\jansky}{Jy}
\DeclareSIUnit{\parsec}{pc}
\DeclareSIUnit{\year}{year}

\received{July 20, 2021}
\submitjournal{ApJ}

\shorttitle{Event Rate of SGRB Afterglows with CHIME}
\shortauthors{Shikauchi \textit{et al.}}

\begin{document}

\title{On the use of CHIME to Detect Long-Duration Radio Transients \\ from Neutron Star Mergers}

\correspondingauthor{Minori Shikauchi}
\email{shikauchi@resceu.s.u-tokyo.ac.jp}

\author{Minori Shikauchi}
\affiliation{Department of Physics, the University of Tokyo,
7-3-1 Hongo, Bunkyo, Tokyo 113-0033, Japan}
\affiliation{Research Center for the Early Universe (RESCEU), the University of Tokyo,
7-3-1 Hongo, Bunkyo, Tokyo 113-0033, Japan}
\author{Kipp Cannon}
\affiliation{Research Center for the Early Universe (RESCEU), the University of Tokyo,
7-3-1 Hongo, Bunkyo, Tokyo 113-0033, Japan}
\author{Haoxiang Lin}
\affiliation{Department of Astronomy, the University of Tokyo,
7-3-1 Hongo, Bunkyo, Tokyo 113-0033, Japan}
\author{Tomonori Totani}
\affiliation{Department of Astronomy, the University of Tokyo,
7-3-1 Hongo, Bunkyo, Tokyo 113-0033, Japan}
\affiliation{Research Center for the Early Universe (RESCEU), the University of Tokyo,
7-3-1 Hongo, Bunkyo, Tokyo 113-0033, Japan}
\author{J. Richard Shaw}
\affiliation{Department of Physics and Astronomy, University of British Columbia, 6224 Agricultural Road, Vancouver, BC V6T 1Z1, Canada}



\begin{abstract}
Short gamma-ray burst (SGRB) GRB 170817A was found to be related to a binary neutron star (BNS) merger.
It is uncertain whether all SGRBs are caused by BNS mergers, and also
under what conditions a BNS merger can cause a SGRB.
As BNS mergers can cause SGRBs, afterglow observations will also provide an alternative measurement of the BNS merger rate independent of gravitational-wave observations.
In previous work by \cite{Feng2014}, the feasibility of the detection of afterglows was considered using a variety of radio observatories and a simple flux threshold detection algorithm.
Here, we consider a more sophisticated detection algorithm for SGRB afterglows, and provide an estimate of the trials factors for a realistic search to obtain an updated estimate of the possibility of observing afterglows with the Canadian Hydrogen Intensity Mapping Experiment (CHIME).
We estimate \SI{\ourrealnc}{ } and \SI{\ourrealc}{ } afterglows per year can be detected using a $3 \sigma$ confidence level threshold with two jet models, one with half opening angle uniformly distributed in $\SI{6}{\degree}$ to $\SI{30}{\degree}$ and the other uniformly distributed in $\SI{3}{\degree}$ to $\SI{8}{\degree}$ with the median $\SI{6}{\degree}$.
We also find \offaxisrationc\% and \offaxisratioc\%, respectively, of the detectable afterglows for each jet opening distribution are off-axis, which are candidates for orphan afterglows.
Our result predicts fewer detectable sources per year than the earlier analysis, but confirms the essential conclusion that using CHIME to search for afterglows will be effective at constraining the astrophysical merger rate.
\end{abstract}


\section{Introduction}
\label{sec:intro}
Gamma-ray bursts (GRBs) are extremely energetic events in the Universe. For short GRBs (SGRBs), prompt emissions are short-duration, intense pulses that last shorter than $\sim\SI{2}{\second}$.
There have been 125 detected with the SWIFT/Burst Alert Telescope\footnote{\url{https://swift.gsfc.nasa.gov/archive/grb_table}} so far.
As time passes, long-lasting afterglows can be observed with durations ranging from months to years, much longer than prompt emissions.
They are very faint ranging from microjanskys to millijanskys.
Among the 125 detections, 83 afterglows were detected in X-ray, 15 in UV/optical and only 6 in radio band. Afterglows in the radio band are so faint that it is difficult to observe them.

GRB~170817A \citep{Abbottetal2017BNSandSGRB} is a SGRB detected with {\it Fermi}-GBM \citep{Goldsteinetal2017} and INTEGRAL \citep{Savchenkoetal2017}, which arrived \SI{1.7}{s} after GW170817 \citep{Abbottetal2017BNS}, a gravitational wave (GW) believed to be from a BNS merger.
It is an unusual SGRB because it is much fainter than is typical \citep{GCN_Connaughtonetal2017,GCN_Goldsteinetal2017,GCN_vonKienlinetal2017}.
In \cite{Kasliwaletal2017}, there are some proposed models explaining fainter SGRBs.
One is a relativistic jet with misaligned observer. Another includes a ``cocoon'', \textit{i.e.}, mildly relativistic matter outflow in addition to the former model. The cocoon is produced by a jet which either successfully breaks out of the BNS merger ejecta or fails (choked jet) \citep{Mooleyetal2018a,Nakaretal2018} or by the so-called fast tail of the dynamical merger ejecta \citep{Mooleyetal2018a,Hotokezakaetal2018b} initially driven by the shock wave formed at the collision front \citep{Bausweinetal2013,Hotokezakaetal2013,Kyutokuetal2013,Kiuchietal2017}.
After the observation of GRB~170817A's prompt emission, follow-up observations from UV to near-infrared wavelength took place and an electromagnetic counterpart SSS17a/AT2017gfo was observed \citep{Arcavietal2017,Chornocketal2017,Coulteretal2017,Cowperthwaiteetal2017,Droutetal2017,Evansetal2017,Kasenetal2017,Kilpatricketal2017,Piranetal2017,Savchenkoetal2017,Shappeeetal2017,Smarttetal2017,Tanaketal2017,Tanviretal2017,Valentietal2017,Villaretal2017}. The host galaxy of the counterpart is identified as NGC~4993, which is an elliptical galaxy with a luminosity distance of $\sim\SI{40}{\mega\parsec}$ \citep{Coulteretal2017}.
X-ray and radio emission was detected 9 and 16~days after the detection of GRB~170817A \citep{Hallinanetal2017,Trojaetal2017}. Based on about one year observation, the brightness changed $\propto t_{\rm{obs}}^{0.8}$ \citep{Resmietal2018} and turned over after around 150~days \citep{Dobieetal2018}. Its energy spectrum follows a single and constant power-law distribution $F_{\nu} \propto \nu^{-0.6}$ consistent with synchrotron radiation \citep{Alexanderetal2018,DAvanzoetal2018,Dobieetal2018,Haggardetal2017,Hallinanetal2017,Lymanetal2018,Marguittietal2017,Marguittietal2018,Mooleyetal2018a,Resmietal2018,Trojaetal2017,Trojaetal2018,vanEertenandHendrik2018}.
However, the observed rising pattern challenges a homogeneous jet model or single-velocity spherical shell model of expanding ejecta because both models generate a faster rise in flux $F_{\nu} \propto t_{\rm{obs}}^3$ \citep{Mooleyetal2018a,NakarandPiran2018}.
Therefore, the homogeneous jet models are excluded and afterglow observations can help reveal the details of the relativistic jet.
There is another important implication from GRB~170817A. Since BNS mergers can cause SGRBs, afterglow observations will provide an alternative
measurement of the BNS merger rate independent of GW observations. 
As there will be contamination from astrophysical phenomena with the similar explosion mechanisms such as long GRBs, we will only obtain an upper bound on the afterglow event rate from afterglow observations.
Though both the BNS merger rate and SGRB event rate have been already measured \citep{Abbottetal2017BNS,Fongetal2015}, the afterglow event rate has never estimated.

The afterglows observed so far were detected in follow-up observations of prompt emissions.
Outside of the jet opening angle, prompt emission is difficult to observe. However, 
as time passes, the jet breaks and synchrotron emission occurs almost isotropically, and an observer outside of the jet opening can observe this afterglow.
An afterglow observed without prompt emission is called an ``orphan afterglow''.
\cite{Rhoads1997} indicated a number of SGRBs should be $\upgamma$-ray faint, which should be observable as orphan afterglows.
In addition, during the $\upgamma$-ray emission phase, relativistic jets are highly beamed with Lorentz factor $\Gamma \sim 100$ \citep{Fenimoreetal1993,WoodsandLoeb1995}. If the $\upgamma$-ray emissions are beamed into a fraction $f_{\rm{b}}$ of the sky, a jet opening angle $\theta_\mathrm{j}$ can be approximated to $1/\Gamma$ and $f_{\rm{b}}$ should be $(1 - \cos \theta_\mathrm{j}) \sim \frac{1}{2} \theta_\mathrm{j}^2 \sim 1/(2 \Gamma^2)$. While the Lorentz factor is $\sim 100$ during the $\upgamma$-ray emission phase, it decreases to order unity when the radio afterglow emissions occur \citep{Waxmanetal1998}.
Therefore, 
a rate of observing off-axis radio afterglows is $4 \pi/f_\mathrm{b} \sim 4 \pi / (1/(2 \times 100^2)) \sim 2.5 \times 10^5$ times larger than that of observing off-axis $\upgamma$-ray emissions 
and it is worth searching for orphan radio afterglows.
Also, the event rate of orphan afterglows depends on the structure of the relativistic jet \citep{TotaniandPanaitescu2002,Nakaretal2002,Rossietal2008}, and its measurement will be helpful in constraining the progenitor of SGRBs. 
Some previous searches tried to observe orphan afterglows with small aperture telescopes
\citep{Grindlay1999,Greineretal2000,Rauetal2006,Malacrinoetal2007,Huangetal2020,Levinsonetal2002,Gal-Yametal2006}, resulting in no confident detections.

The Canadian Hydrogen Intensity Mapping Experiment (CHIME) will greatly contribute to orphan afterglow searches with its wide instantaneous field of view of $\sim 200$ deg$^2$. A previous work \citep{Feng2014} theoretically estimated $\sim 30$ to 3000 orphan afterglows per year could be detected.
CHIME is a cylindrical transit radio telescope located in British Columbia, Canada. It has four cylinders, each with 256 $\times$ 4 dual-polarization feeds and observes from $\SI{400}{\mega\hertz}$ to $\SI{800}{\mega\hertz}$. At mid-band CHIME has an angular resolution of $\SI{0.22}{\degree}$--$\SI{5.1}{\degree} \times \SI{0.36}{\degree}$ depending on the declination (from projection effects the resolution is lower towards the local horizon).
It also observes the whole Northern sky one per day. 
Therefore, CHIME is expected to produce daily skymaps everywhere in the northern sky and we expect that we would use them for searching for the afterglows.
The CHIME simulation and data analysis pipeline is publicly available \citep{radio_cosmo, chime_exp}.
With these you can simulate the response of the telescope to a simulated skymap.

In the event rate estimation of \cite{Feng2014}, they prepared a set of afterglow light curves using typical isotropic energies and circum medium densities.
They assumed a homogeneous jet model with a fixed jet angle of $\SI{11.5}{\degree}$ ($\SI{0.2}{\radian}$), which is a typical value, and the observer angle changes from $\SI{0}{\degree}$ to $\SI{90}{\degree}$, on-axis to off-axis.
However, the observation of GRB~170817A, the only event known to be related to a BNS merger, implied the existence of more complicated jet models.
As our target is long-duration transients from BNS mergers, we need to prepare light curve templates for complex jet SGRBs.
To that end, \cite{Linetal2019} developed an excellent analytic calculation code for light curves from a relativistic jet with Gaussian energy profile, called ``Gaussian jet model'', and (quasi-)spherical outflow with radially stratified velocity.
The code calculates synchrotron emission from a relativistic blast wave, which is powered by a structured jet viewed from a given angle and propagating in a constant density medium. 
In contrast to the previous efforts 
\citep{DAvanzoetal2018,GillandGranot2018,Hotokezakaetal2018b,Marguittietal2018,Mooleyetal2018a,Nakaretal2018,Resmietal2018,Trojaetal2018}
, \cite{Linetal2019} treated $f$, the number fraction of electrons injected to the shock acceleration process, and $\upgamma_m$, the minimum Lorentz factor of electrons in the shock frame as free parameters.
With few data points, the assumption that all the electrons in the shocked shell should be accelerated, that is $f=1$, was adopted in the previous work, but this might be an oversimplification.
It is more natural to think that some fraction of electrons remain as thermal particles as observed in supernova remnants \citep{Laming2001,Bambaetal2003}.
This treatment is effective if there are many light curve data points in a wide wavelength band.
In addition to that they added some corrections and improvements to treat the radially stratified spherical model carefully.
Therefore, we employ the latest calculation code \citep{Linetal2019} for light curve templates reflecting the properties of GRB~170817A.

In order to discuss the detectability of SGRB afterglows, \cite{Feng2014} set two detection criteria for the prepared light curves. First, the peak flux of a detectable light curve must be larger than
a threshold value. Second, the brightness of a detectable light curve must change in a time scale smaller than the operation time of the radio telescope. 
However, the event rates under the detection criteria might be overestimated because it might be hard to consider the detectability from such a clear light curve. 
Here we focus on the likelihood ratio statistic employed in \cite{Feng2017}.
They employed the statistic to search for long-duration radio transients with the Murchison Widefield Array and also showed that under certain simplifying assumptions the likelihood ratio reduces to a matched filter-like inner product. 
In general, the sensitivity of a source-finding algorithm applied to each image is limited by non-thermal noise sources such as classical confusion noise caused by faint and unresolved background sources.
A simple way to find time-dependent sources is subtracting images taken at the same local sidereal time.
Sidelobe confusion noise caused by residual synthesized beam sidelobes as well as a classical confusion noise can be subtracted.
However, for most surveys, images are not taken at the same sidereal time each day. Sidelobe confusion noise is not negligible and the image subtraction can be dominated by such artifacts.
These artifacts should be treated carefully because they can be identified as astrophysical transients (e.g. \cite{Frailetal2012} reported transient candidates \cite{Boweretal2007} found were artifacts or caused by calibration errors). 
The likelihood ratio statistic cannot reveal sources masked by such noise but can avoid being fooled by it by including knowledge of its statistical properties including the distribution of the artifacts.
In order to estimate the trials factor and set a detection threshold, \cite{Feng2017} processed off-source data and modeled the distribution of observed likelihood ratios by assuming the negative log-likelihood ratio can be treated as an exponential function.
Note that their survey is not dedicated to SGRB afterglows, but for general long-duration transients.
They used light curve templates with a top-hat shape and a fast rise and an exponential decay.

In this work, we adopt the likelihood ratio statistic used in \cite{Feng2017} and develop an analytic estimate of a trials factor in section \ref{sec:method}. We show the results in section \ref{sec:result} and discuss how the choice of physical parameters of a relativistic jet affects the event rate estimation and the possibility of direct detections of orphan afterglows in section \ref{sec:discussion}. 

\section{Method}
\label{sec:method}
The detection algorithm used here is based on the likelihood ratio test. 
First, consider the likelihood ratio $\Lambda$,
\begin{equation}
	\Lambda = \frac{P({\rm{data}}|{\rm{signal}})}{P({\rm{data}}|{\rm{noise}})},
\end{equation}
where $P({\rm{data}}|{\rm{signal}})$ is the probability of obtaining a data set when a signal is present, and $P({\rm{data}}|{\rm{noise}})$ is that of obtaining a data set when no signals are included.

By generating mock data including only noise and calculating $\Lambda$, we can obtain $P(\ln \Lambda|{\rm{noise}})$, the probability of obtaining $\ln \Lambda$ given that only noise is present.
Note that $P(\ln \Lambda|{\rm{noise}})$ is the probability of $\ln \Lambda$ for one pixel being compared to one light curve template.
Considering that CHIME's data is multi-pixel and we employed a suite of light curve templates, we can set a threshold value $\Lambda_{\rm{th}}$ satisfying below,
\begin{equation}
	P_{\rm{exp}}(\ln \Lambda \geq \ln \Lambda_{\rm{th}}| {\rm{noise}}) < p,
\end{equation}
where $P_{\rm{exp}}(\ln \Lambda \geq \ln \Lambda_{\rm{th}}|{\rm{noise}})$ is the probability in ``one experiment'' of obtaining $\Lambda$ larger than the threshold value $\Lambda_{\rm{th}}$ given only noise is present, and $p$ is the ``false alarm probability'' (FAP).
The meaning of ``one experiment'' is the full visible sky for the full duration of the observation and the full suite of light curve templates. We assume a one year observation at the center of the frequency band, $\SI{600}{\mega\hertz}$, for all the pixels.
If we employ a longer time, the number of detectable events will increase.

As we will show later, $\Lambda$ can be approximated like an inner product between data and a light curve template. 
The value $P_{\rm{exp}}(\ln \Lambda \geq \ln \Lambda_{\rm{th}}|{\rm{noise}})$ can be calculated by subtracting the probability of obtaining $\ln \Lambda$ smaller than a threshold value $\ln \Lambda_\mathrm{th}$ for all the pixels and statistically independent combinations of light curve templates,
\begin{equation}
	P_{\rm{exp}}(\ln \Lambda \geq \ln \Lambda_{\rm{th}}|{\rm{noise}}) = 1 - (1 - P(\ln \Lambda \geq \ln \Lambda_{\rm{th}}|{\rm{noise}}))^n,
\end{equation}
where 
\begin{equation}
	n = n_{\rm{pix}} \times n_{\rm{dof}}, 
\end{equation}
$n_{\rm{pix}}$ is the number of independent pixels in an output image, $511 \times 1409$ and $n_{\rm{dof}}$ is the number of degrees of freedom (DOF) of the light curve templates, the number of statistically independent light curve templates, which we find to be $5$. 
The value of $n_{\rm{dof}}$ is estimated from a singular-value decomposition (SVD), a matrix factorization that expands a matrix as a product of
two orthogonal matrices and a diagonal matrix. By populating a matrix
with many random light curve templates, decomposing it
using the SVD, and counting the number of non-zero elements in the
diagonal matrix in the factorization, we see the number of orthogonal
vectors that are needed to reconstruct any of the light curve
templates. 
For the SVD, we generated $\sim 100,000$ light curve fragments with isotropic kinetic energy $\SIlist{1e49;1e51}{\erg}$, circum medium density $\SIlist{1e-5;1e-3;1}{\centi\metre^{-3}}$, observer angle distributed uniformly on the sphere between $0$ and $1$, with a fixed jet opening angle $\SI{0.2}{\radian}$.
For more detail, refer to Appendix \ref{app:svd}. 

For each pixel of the sky map, we assume the data set $x_i$ ($i=1, 2, \cdots ,365$) is a time-series of ``pixel brightness'', which is obtained by converting interferometer visibilities into a flux in millijansky.
We consider two sources of noise: a thermal noise component $\sigma_i$ arising in the antenna and electronics, and a background component $c_i$ from celestial radio sources.
If the data contains only noise (hypothesis 1) then
\begin{equation}
	x_i = \sigma_{i} + c_i,
\end{equation}
and when a source is present (hypothesis 2)
\begin{equation}
	x_i = \sigma_{i} + c_i + A f_i,
\end{equation}
where $f_i$ is a light curve template we prepared and $A$ is its amplitude.
For simplicity we ignore time-dependence in the thermal noise and celestial noise and celestial background. The net, constant, noise process leads to a variance in the observed flux of $\sigma^2 = (\SI{0.0232}{\milli\jansky})^2$.
The derivation of the variance is shown in Appendix \ref{app:noise_est}.
Since a typical value of the constant background is $\sim\SI{10}{\kelvin}$, corresponding to $\SI{2.76}{\jansky\per\text{pixel}}$, we employed $\SI{2.76}{\jansky}$ as the average value of $c_i$, $<c_i> \equiv c$.

For hypothesis 1, we assume white stationary noise, and the probability density function for the pixel brightness $x_i$, $p_i(x_i)$ is a Gaussian distribution,
\begin{equation}
	p_i(x_i) = \frac{1}{\sqrt{2 \pi} \sigma} \exp \left [ -\frac{(x_i - c)^2}{2 \sigma^2} \right ].
\end{equation}
Therefore, the numerator of the likelihood ratio $P({\rm{data}}|{\rm{noise}})$ can be expressed as
\begin{equation}
	P({\rm{data}}|{\rm{noise}}) = {\prod_{i=1}^{365}} p_i(x_i).
\end{equation}

Also, for hypothesis 2, the probability density function of a pixel brightness $p_i'(x_i)$ will change by a factor of $Af_i$,
\begin{equation}
	p'_i(x_i) = \frac{1}{\sqrt{2 \pi} \sigma} \exp \left [ -\frac{(x_i - c - Af_i)^2}{2 \sigma^2} \right ],
\end{equation}
and we obtain $P({\rm{data}}|{\rm{signal}})$,
\begin{equation}
	P({\rm{data}}|{\rm{signal}}) = {\prod_{i=1}^{365}} p'_i(x_i).
\end{equation}

Now, we have two parameters, the constant background $c$ and the signal amplitude $A$.
We extremize the likelihood ratio with respect to these parameters.
We have only to consider the exponents of $P({\rm{data}}|{\rm{noise}})$ and $P({\rm{data}}|{\rm{signal}})$, defined as $\chi_1^2$ and $\chi_2^2$,
\begin{align}
	\chi^2_1 &= \sum_{i=1}^{365} \frac{(x_i - c)^2}{\sigma^2}, \\
	\chi^2_2 &= \sum_{i=1}^{365} \frac{(x_i - c - Af_i)^2}{\sigma^2},
\end{align}
following the previous work \citep{Feng2017}.  We start to solve for the extrema with
\begin{equation}
	\frac{\partial \chi^2_1}{\partial c} = - \sum_{i=1}^{365} \frac{2 (x_i - c)}{\sigma^2}.
\end{equation}
For $\frac{\partial \chi^2_1}{\partial c} = 0$, we obtain
\begin{equation}
	c_1 = \frac{\sum \frac{x_i}{\sigma^2}}{\sum \frac{1}{\sigma^2}} \equiv \langle \bm{x} \rangle,
\end{equation}
\begin{equation}
	\frac{\partial \chi^2_2}{\partial c} = - \sum_{i=1}^{365} \frac{2 (x_i - c - Af_i)}{\sigma^2}, \;
	\frac{\partial \chi^2_2}{\partial A} = - \sum_{i=1}^{365} \frac{2 f_i (x_i - c - Af_i)}{\sigma^2}.
\end{equation}
For $\frac{\partial \chi^2_2}{\partial c}, \frac{\partial \chi^2_2}{\partial A} = 0$, 
\begin{eqnarray}
	\begin{cases}
		{\displaystyle{\sum_{i=1}^{365} \frac{1}{\sigma^2}}} \langle \bm{f} \rangle A_2 + {\displaystyle{\sum_{i=1}^{365} \frac{1}{\sigma^2}}} c_2 = {\displaystyle{\sum_{i=1}^{365} \frac{1}{\sigma^2}}} \langle \bm{x} \rangle & \\
		(\bm{f}, \bm{f}) A_2 + {\displaystyle{\sum_{i=1}^{365} \frac{1}{\sigma^2}}} \langle \bm{f} \rangle c_2 = (\bm{x}, \bm{f}), &
	\end{cases}
\end{eqnarray}
is derived where
\begin{equation}
	\langle \bm{x} \rangle \equiv \frac{\sum \frac{x_i}{\sigma^2}}{\sum \frac{1}{\sigma^2}}, \;
	(\bm{x}, \bm{f}) \equiv \sum_{i = 1}^{L} \frac{x_i f_i}{\sigma^2}.
\end{equation}
Then, $c_2$ and $A_2$, which extremize $\chi_2^2$, are expressed as
\begin{align}
	c_2 &= \langle \bm{x} \rangle - A_2 \langle \bm{f} \rangle \\
	A_2 &= \frac{(\bm{x}, \bm{f}) - {\displaystyle{\sum_{i=1}^{365} \frac{1}{\sigma^2}}} \langle \bm{x} \rangle \langle \bm{f} \rangle}{(\bm{f}, \bm{f}) - {\displaystyle{\sum_{i=1}^{365} \frac{1}{\sigma^2}}}\langle \bm{f} \rangle^2} = \frac{(\bm{x}, \bm{f} - \langle \bm{f} \rangle)}{(\bm{f} - \langle \bm{f} \rangle, \bm{f} - \langle \bm{f} \rangle)}.
\end{align}
By substituting $c_1, c_2$ and $A_2$ into the likelihood ratio, we obtain the maximum likelihood ratio $\Lambda_{\rm{max}}$,
\begin{align}
	\ln \Lambda_{\rm{max}}&= \frac{1}{2} \times \frac{(\bm{x}, \bm{f} - \langle \bm{f} \rangle)^2}{(\bm{f} - \langle \bm{f} \rangle, \bm{f} - \langle \bm{f} \rangle)},
\label{eq_max_lnL}
\end{align}
which is defined as $\rho^2/2 \sigma_{\rho}^2$ in \cite{Feng2017}.

In order to estimate an event rate of SGRB afterglows, we prepared 30,000 light curve templates. 
The energy and Lorentz factor distributions are assumed to be angle-dependent from the symmetric axis and Gaussian distributions, called ``Gaussian jet model''. 
The intrinsic distribution of isotropic kinetic energy and circum medium density follows \citep{Fongetal2015} based on afterglow observations from prompt emissions of SGRBs.
We employed their distributions of isotropic kinetic energy $E_\mathrm{k, iso}$ and circumburst density $n$ with a fixed choice of $\epsilon_e = \epsilon_B = 0.1$. While lower values of $\epsilon_B$ are supported in some literature, we note that choosing a smaller $\epsilon_B$ requires increasing the energy and density correspondingly in order to explain the brightness of the same observed afterglow data in \cite{Fongetal2015}. For example, if we input $\epsilon_B = 10^{-4}$ instead, then the energy and density should be increased by factors of 10 compared to the $\epsilon_B = 0.1$ case, as suggested by \cite{Fongetal2015}. As a result, the flux densities predicted by our model in the range of interest will only change by a factor of order unity. Hence, we do not expect a significant change in the predicted event rate by different choices of parameter distributions. 

For a jet opening angle, \cite{Fongetal2015} estimated three intrinsic distributions for SGRBs. 
Among them, we employed two different distributions and call them ``Model A'' and ``Model B'' in this paper. 
Model A is based on 11 SGRBs with jet break detected or lower limit placed, with an ad hoc upper limit at $\ang{30;;}$ and Model B is based on only 4 SGRBs with jet break detected.
The measured jet opening angles are in $\ang{3;;}$ -- $\ang{8;;}$ and the median is $\ang{6;;} \pm \ang{1;;}$ .
Note that the rest one has an ad hoc upper limit at $\ang{90;;}$ and seems unphysical because the reasonable range of a jet opening angle is $\ang{5;;}$ to $\ang{30;;}$ based on numerical simulations of post-merger black hole accretion \citep{RuffertandJanka1999, RosswogandRamirezRuiz2003, Alloyetal2005, Rosswog2005, Rezzollaetal2011}. 
We distributed $195,000$ simulated radio signals into space and did a Monte-Carlo simulation to obtain the fraction of detectable signals at a given comoving distance $r$, $P({\rm{detection}}|r)$.
The number of distributed signals is proportional to comoving distance squared up to $\SI{1500}{\mega\parsec}$. 
The real astrophysical cutoff of CHIME is unknown and the event rate depends on it. 
However, the distance cut-off in this work is large enough to discuss the event rate 
because the fraction of detectable signals at such a far distance is very small. 
We assume the direction to the observer is uniformly distributed on the sphere. 

The value $P({\rm{detection}}|r)$ can be expressed as 
\begin{equation}
    P(\mathrm{detection}|r) = P(\ln \Lambda \geq \ln \Lambda_{\mathrm{th}}|\mathrm{signal}) \times P(\mathrm{not \;obscured}),
\end{equation}
where $P(\mathrm{not \;obscured})$ is the probability that a signal is not obscured by other, closer, brighter, sources. 
As the number of detectable sources increases, a signal is more likely to be obscured by other sources.
The probability a signal is not obscured by other sources is
\begin{equation}
    P(\mathrm{not \;obscured}) = \left ( \frac{n_{\mathrm{pix}} - 1}{n_{\mathrm{pix}}} \right )^{(\text{\# of closer sources})},
\end{equation}
where $(n_{\mathrm{pix}} - 1)/n_{\mathrm{pix}}$ is the probability that one of the closer sources is not occupying the same pixel.

Assuming the progenitors of SGRBs are BNS mergers, the event rate of detectable SGRB afterglows $N_{\rm{det}}$ can be calculated by integrating $r^2 \times P({\rm{detection}}|r)$ up to 1500 Mpc,
\begin{equation}
	N_{\rm{det}} = R_{\rm{BNS}} \times \frac{1}{2} \times 4 \pi \int r^2 P({\rm{detection}}|r) \mathrm{d}r.
\label{eq_event_rate}
\end{equation}
The BNS merger rate from GW observations is \citep{Abbottetal2020}
\begin{align}
	R_{\rm{BNS}} = 980^{+1490}_{-730} \; {\rm{Gpc}}^{-3} \; {\rm{year}}^{-1}\\
	{\rm{upper \; limit }}\; : \; 12600 \; {\rm{Gpc}}^{-3} \; {\rm{year}}^{-1}.
\end{align}
Note that the factor $1/2$ in \eqref{eq_event_rate} is based on the fact that CHIME observes all the Northern sky.

\section{Result}
\label{sec:result}
\subsection{Selection of Detection Threshold}
The detection threshold must be relatively high because of the large number of pixels and light curves we consider, but because end-to-end simulations including the construction of synthetic data sets and the analyses of those data sets are costly, it is difficult to obtain, via brute-force, enough samples of $\ln \Lambda$ to measure the ranking statistic's distribution to sufficiently small false-alarm probabilities to choose a suitable detection threshold.  Therefore, we begin by obtaining an analytic estimate of the distribution.
\begin{figure}
	\centering
	\includegraphics[width=\width]{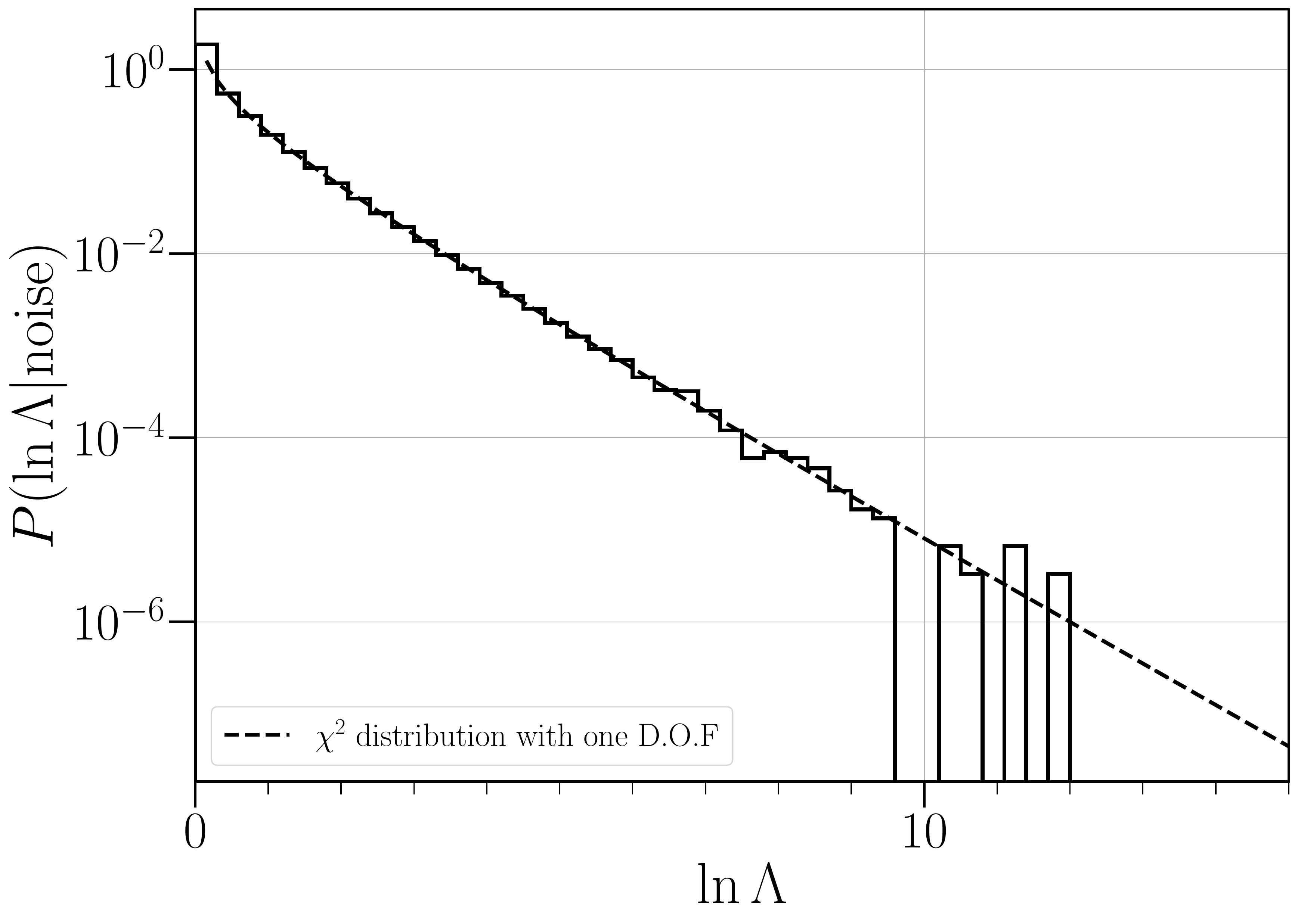}
	\caption{Distribution of $P(\ln \Lambda | {\rm{noise}})$ in one pixel and one light curve template. It is obtained by generating random noise data and calculating $\ln \Lambda$. From Wilks' theorem \citep{wilks1938}, it should follow a $\chi^2$ distribution and we confirmed a $\chi^2$ distribution with one DOF, the dashed line, matched our result. We employed the $\chi^2$ distribution in order to calculate FAP in one experiment.}
	\label{wilks_lnL}
\end{figure}
In Figure 1, the solid line is a histogram of outcomes after 1,000,000 random simulation showing the distribution of $\ln \Lambda$ under hypothesis 1: only noise are included in the data set of one pixel and one template, $P (\ln \Lambda|\mathrm{noise})$. Where the expected count of outcomes per bin is close to 1 (probabilities near $10^{-6}$ because of the number of samples drawn) the histogram becomes noisy, which, here, occurs near $\ln \Lambda$ of 10.  We require a model of the distribution well beyond this. 
According to Wilks' theorem \citep{wilks1938}, $2 \ln \Lambda$ under hypothesis 1 should follow a $\chi^2$ distribution with some number of DOF. 
Since the number of fitting parameters for $P({\rm{data}}|{\rm{signal}})$ and $P({\rm{data}}|{\rm{noise}})$ is $2$ and $1$ respectively, the likelihood ratio should follow a $\chi^2$ distribution with one DOF. 
The dashed line in Figure \ref{wilks_lnL} shows a scaled $\chi^2$ distribution with one DOF and we see good agreement with the observed distribution of $\ln \Lambda$.  To proceed, we adopt this $\chi^2$ distribution to model the distribution of $\ln \Lambda$ under the noise hypothesis.

\begin{figure}
	\centering
	\plottwo{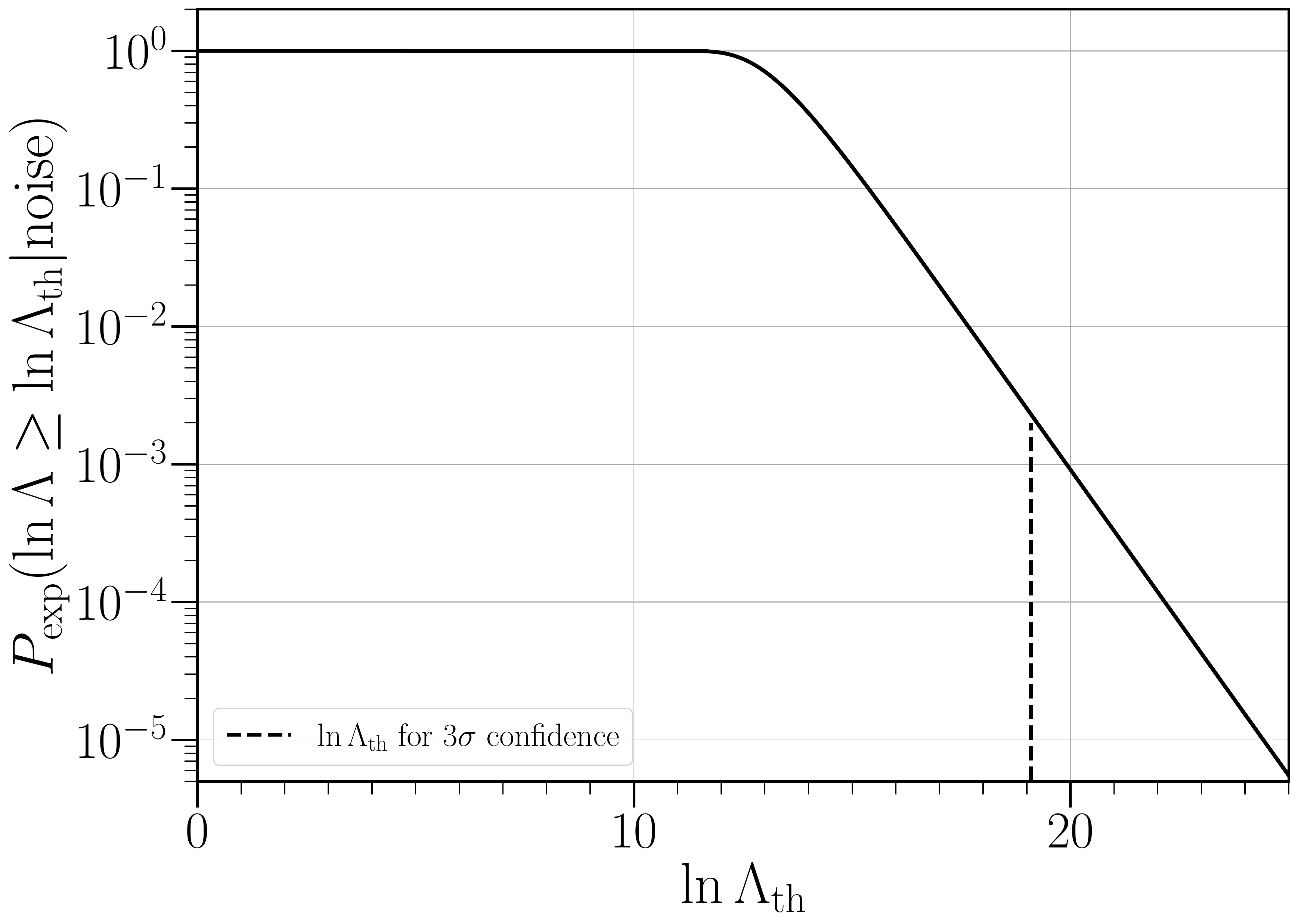}{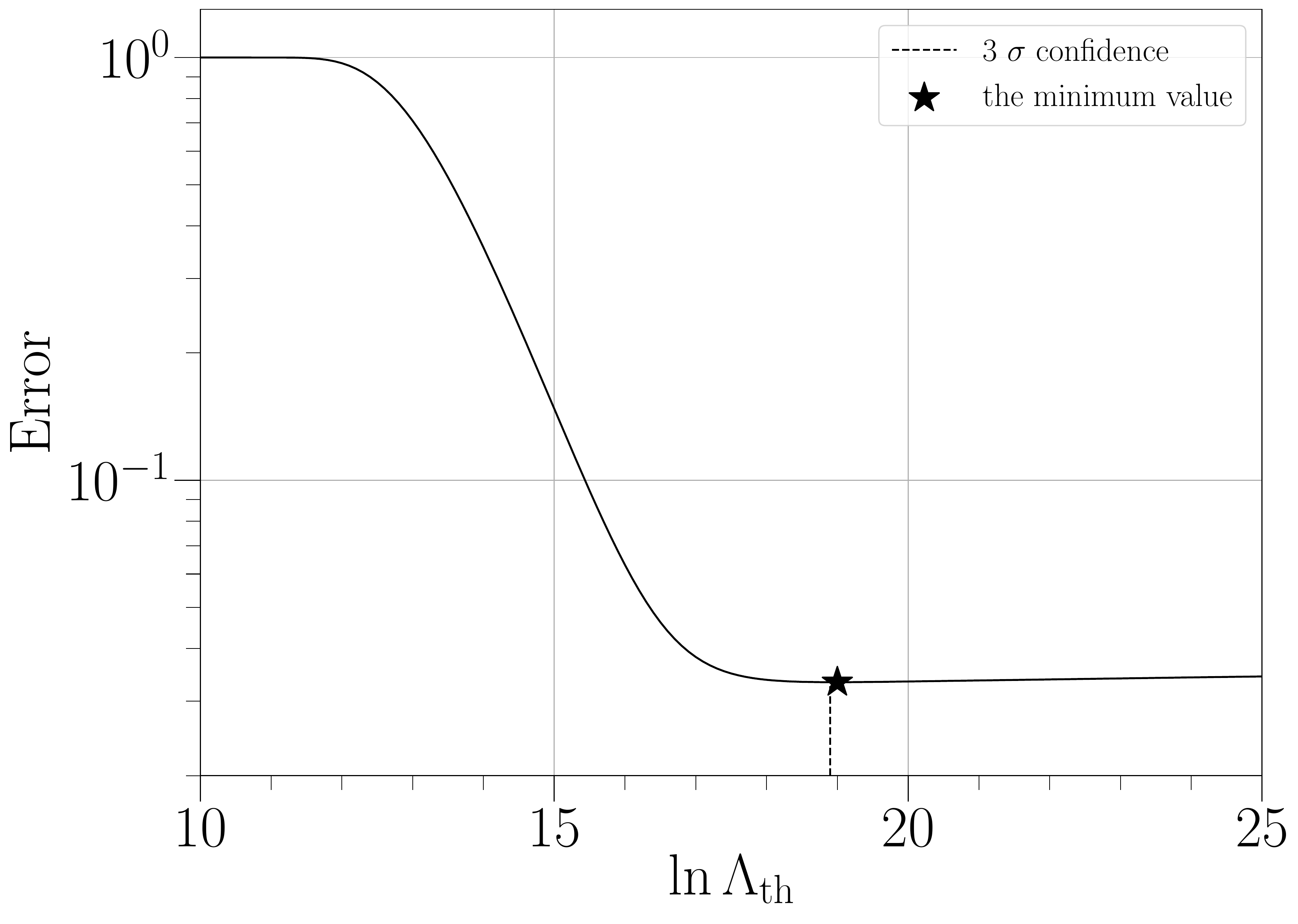}
	\caption{(Left) Probability of obtaining $\ln \Lambda \geq \ln \Lambda_{\rm{th}}$ in one experiment given a data set containing only noise. The threshold corresponding to $3\sigma$ confidence level is shown as a dashed line at $\ln \Lambda_\mathrm{th} = 19.1$. 
	(Right) The distribution of an error in the number of expected events as a function of a likelihood ratio $\ln \Lambda$. The minimum value (a star marker) is $\sim 19$, very close to the $3 \sigma$ confidence level. Therefore, we employed a threshold corresponding to $3 \sigma$ confidence level.}
	\label{lnL_exp_chi_w_line}
\end{figure}

Using the $\chi^2$ distribution, we calculated the probability of obtaining $\ln \Lambda$ larger than a threshold value $\ln \Lambda_{\rm{th}}$ under hypothesis 1, $P_{\rm{exp}}(\ln \Lambda \geq \ln \Lambda_{\rm{th}}|{\rm{noise}})$ (\textit{i.e.}, the FAP), which is shown in the left panel of Figure \ref{lnL_exp_chi_w_line}.  As the threshold is lowered the event rate estimate will come to be in error due to some unknown Poisson-distributed number of false positives, however as the threshold is raised the event rate estimate will come to be in error due to the variance in the small Poisson-distributed number of detected signals.  We select a threshold that minimizes the combination of these two effects.
In the right panel of Figure \ref{lnL_exp_chi_w_line}, the expected fractional error in the estimated rate has a minimum around $\ln \Lambda = 19$. From the left panel, this value corresponds to approximately a $3 \sigma$ confidence level.
Therefore, we employ a detection threshold corresponding to $3 \sigma$ confidence to discuss event rate estimation and direct detection. 

\subsection{Event Rate Estimation}
\begin{figure}
	\centering
	\plottwo{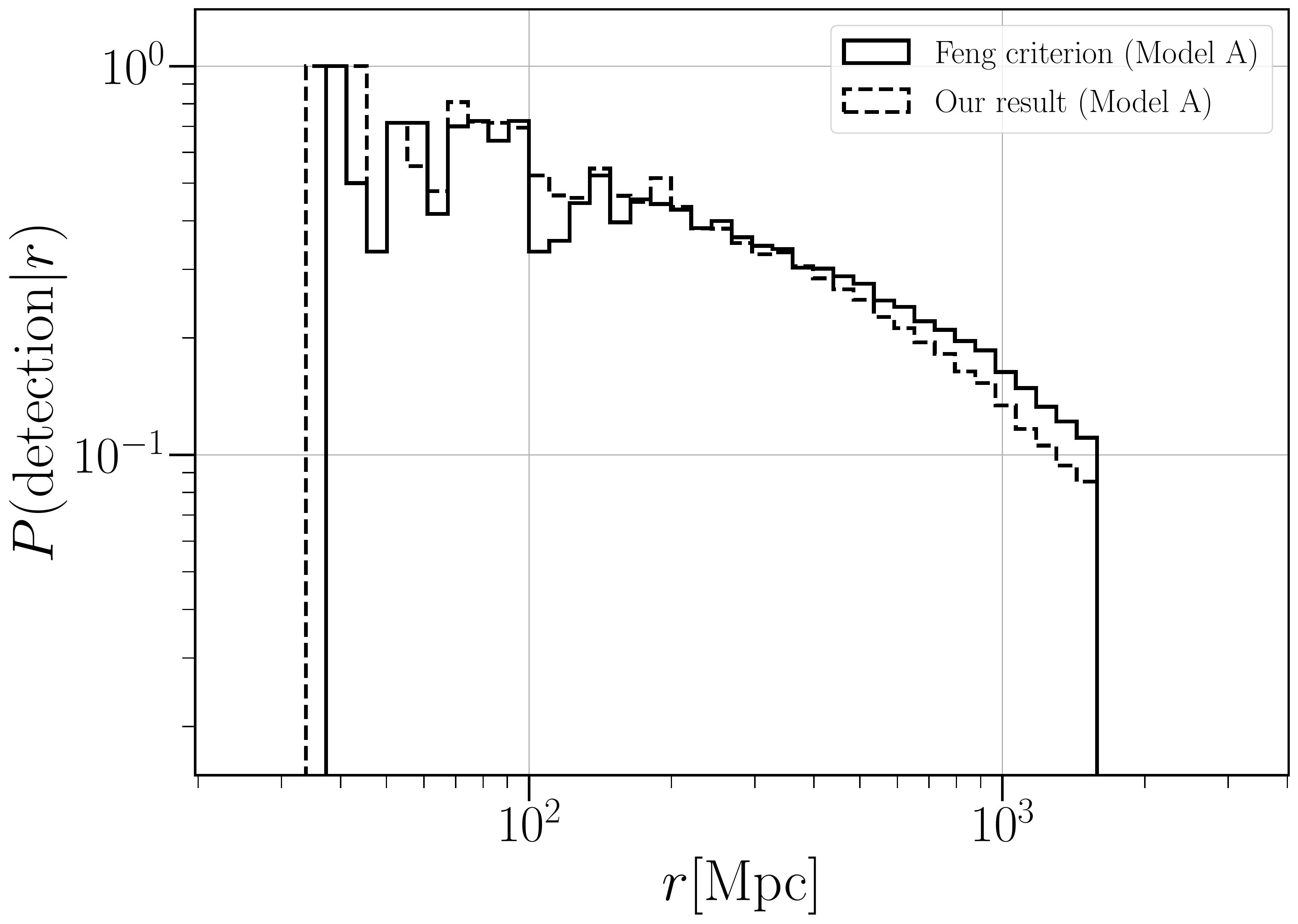}{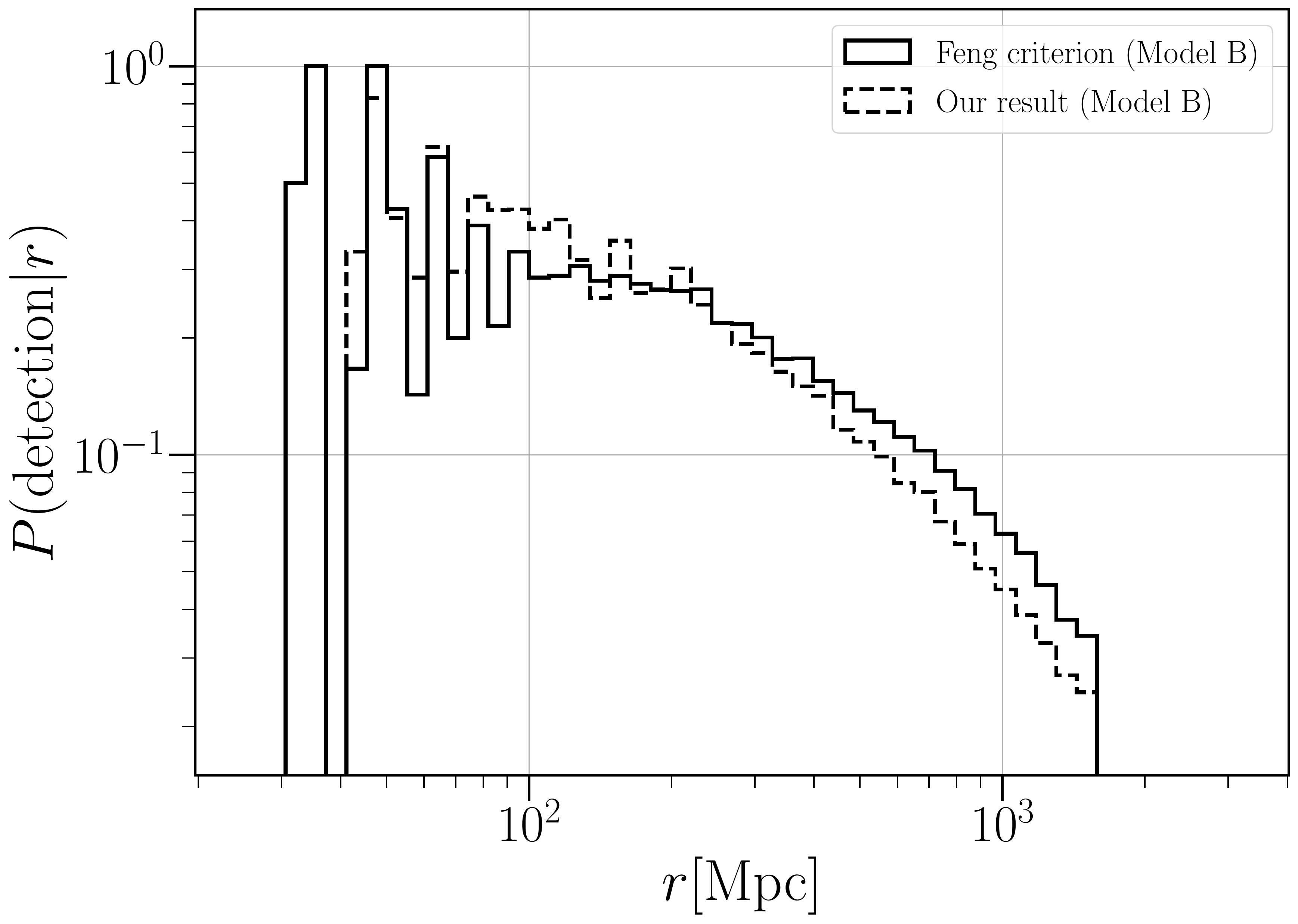}
	\caption{(Left) The probability that a signal located at some comoving distance can be detected with Model A, a jet opening angle distribution with half opening angle uniformly distributed in $6$ -- $30$ deg. The dashed line histogram is based on the detection algorithm and solid one is based on the algorithm the previous work \cite{Feng2014} employed. By integrating the fractions, we obtained $\SI{\ourrealnc}{\per\year}$ and $\SI{\Fengrealnc}{\per\year}$ SGRB afterglows can be detected with our criterion and Feng criterion respectively. (Right) The same as the left panel but using Model B, the jet opening angle distribution peaked at $6$ deg. We obtained $\SI{\ourrealc}{\per\year}$ and $\SI{\Fengrealc}{\per\year}$ afterglows are detectable.}
	\label{lum_vs_prob}
\end{figure}

The left and right panels in Figure \ref{lum_vs_prob} show the fraction of detectable signals at a fixed comoving distance $r$, $p({\rm{detection}}|r)$ with different jet opening angle models, Model A and B, respectively.
In each panel, the dashed line histogram shows the distribution based on our detection criterion.
For the comparison, we calculated the fraction using detection criteria employed in the previous work \citep{Feng2014}, the solid line histogram called ``Feng criterion''.
For both panels, the difference between our criterion and Feng criterion is greater at farther distances.  Since the number density of signals increases in proportion to $r^2$, although the difference appears small, it has a noticeable effect on the estimated event rate.
From the Monte Carlo simulation, we estimated $\SI{\ourrealnc}{\per\year}$ SGRB afterglows can be detected with Model A. In Model B, $\SI{\ourrealc}{\per\year}$ are detectable.  
The detectability with Model B decreases because Model B has a small median value and afterglows become fainter.
Note that the Monte Carlo simulation distributes simulated sources to minimize the sampling noise in the rate estimate, not to minimize the bin-count noise in these histograms:  the sparse samples at small distances is attention getting in the histograms, but those bins contribute a negligible volume to the simulation and a similarly negligible amount of sampling noise to the final rate estimate.
Since the histogram is sparse at closer distance, we interpolate it so that the fraction of detectable signals at $\SI{0}{\mega\parsec}$ is unity. This interpolation does not matter for the event rate estimation.
We summarized the event rates with different detection criteria and jet angle distributions in Table \ref{table_eventrate}.
The first line shows the result of \cite{Feng2014} for a comparison. Note that they estimated the event rate with respect to circum medium density.
The second and third lines are the event rates based on Feng criterion with Model A and B respectively. 
The fourth and fifth lines show those based on our detection criterion with different jet angle distributions.
By comparing the first line to the second and third ones, the event rates are comparable.
The third and fifth lines, our result, are slightly smaller than the rest of the lines.
However, we confirmed the essential conclusion of the previous work, namely that, using CHIME to search for SGRB afterglows is effective.
\begin{table}
\begin{tabular}{cl|cccc} \hline
	& & \multicolumn{4}{c}{Event rate [$\mathrm{year}^{-1}$]} \\ 
	 & & Pessimistic & Realistic & Optimistic & Upper limit  \\ \hline
	\multicolumn{2}{c|}{\cite{Feng2014}}  & 8 -- 750 & 29 -- 2940 & 86 -- 7410 & 504 -- 25200 \\ \hline
	Feng criterion & (model A) & $229$ & $1086$ & $2703$ & $11407$ \\
	$\cdots$ & (model B)       & $107$ & $418$  & $1043$ & $4952$ \\
	Our result & (model A)     & $279$ & $893$  & $2225$ & $10504$ \\ 
	$\cdots$ & (model B)       & $80$  & $312$  & $779$  & $3710$ \\ \hline
\end{tabular}
\caption{Event rate of SGRB afterglows. We estimated $\SI{\ourrealnc}{\per\year}$ and $\SI{\ourrealc}{\per\year}$ afterglows can be detected with Model A and B respectively. Our result estimated slightly smaller value than the previous work \citep{Feng2014} estimated, but confirmed the essential conclusion of \cite{Feng2014}: using CHIME to search for SGRB afterglows will be effective at constraining BNS merger rate.}
\label{table_eventrate}
\end{table}

\subsection{Parameter Distribution of Detectable SGRB Afterglows}
\begin{figure}
\centering
	\includegraphics[width=\width]{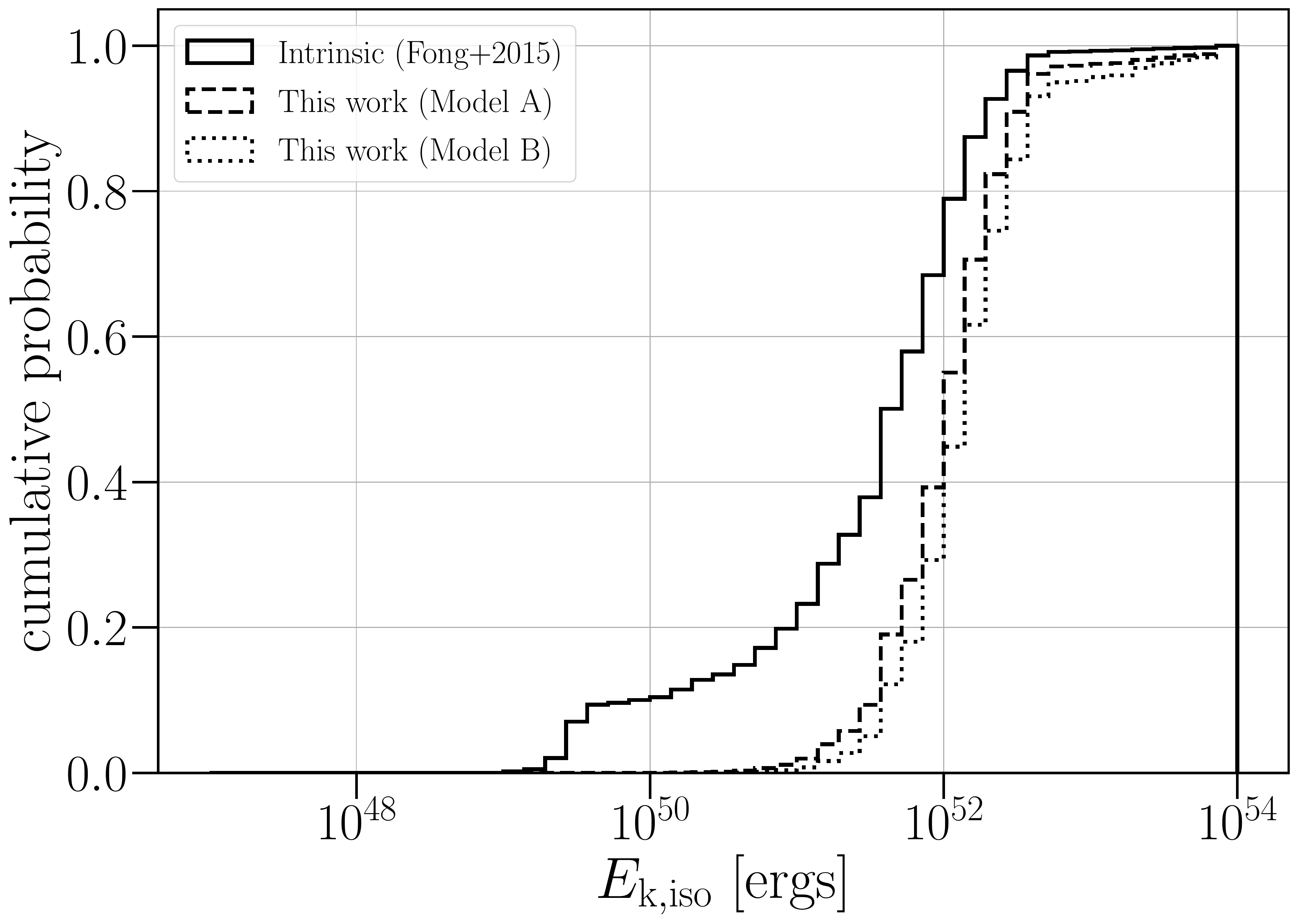}
\caption{Cumulative density function of isotropic kinetic energy of detectable SGRBs. 
	The distributions of detectable signals with Model A and B (dashed and dotted histograms, respectively) are biased to larger values than the intrinsic distribution (the solid one) based on \cite{Fongetal2015}.}
\label{cum_prob_kiso}
\end{figure}

\begin{figure}
\centering
	\includegraphics[width=\width]{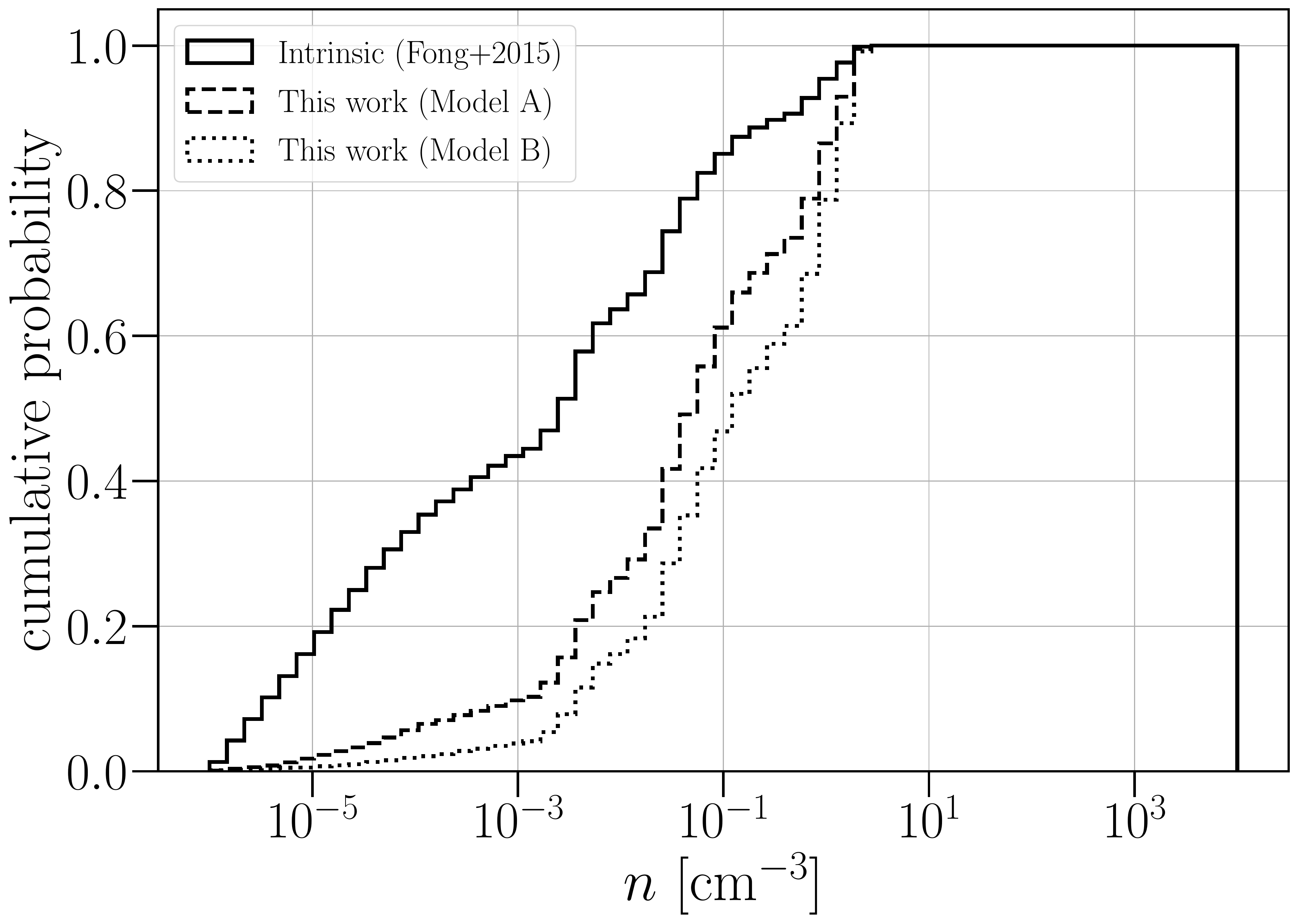}
	\caption{Cumulative density function of isotropic kinetic energy of detectable SGRBs. 
	The distributions of detectable signals with Model A and B (dashed and dotted histograms, respectively) are biased to larger values than the intrinsic distribution (the solid one) based on \cite{Fongetal2015}.}
\label{cum_prob_n}
\end{figure}

\begin{figure}
\centering
	\plottwo{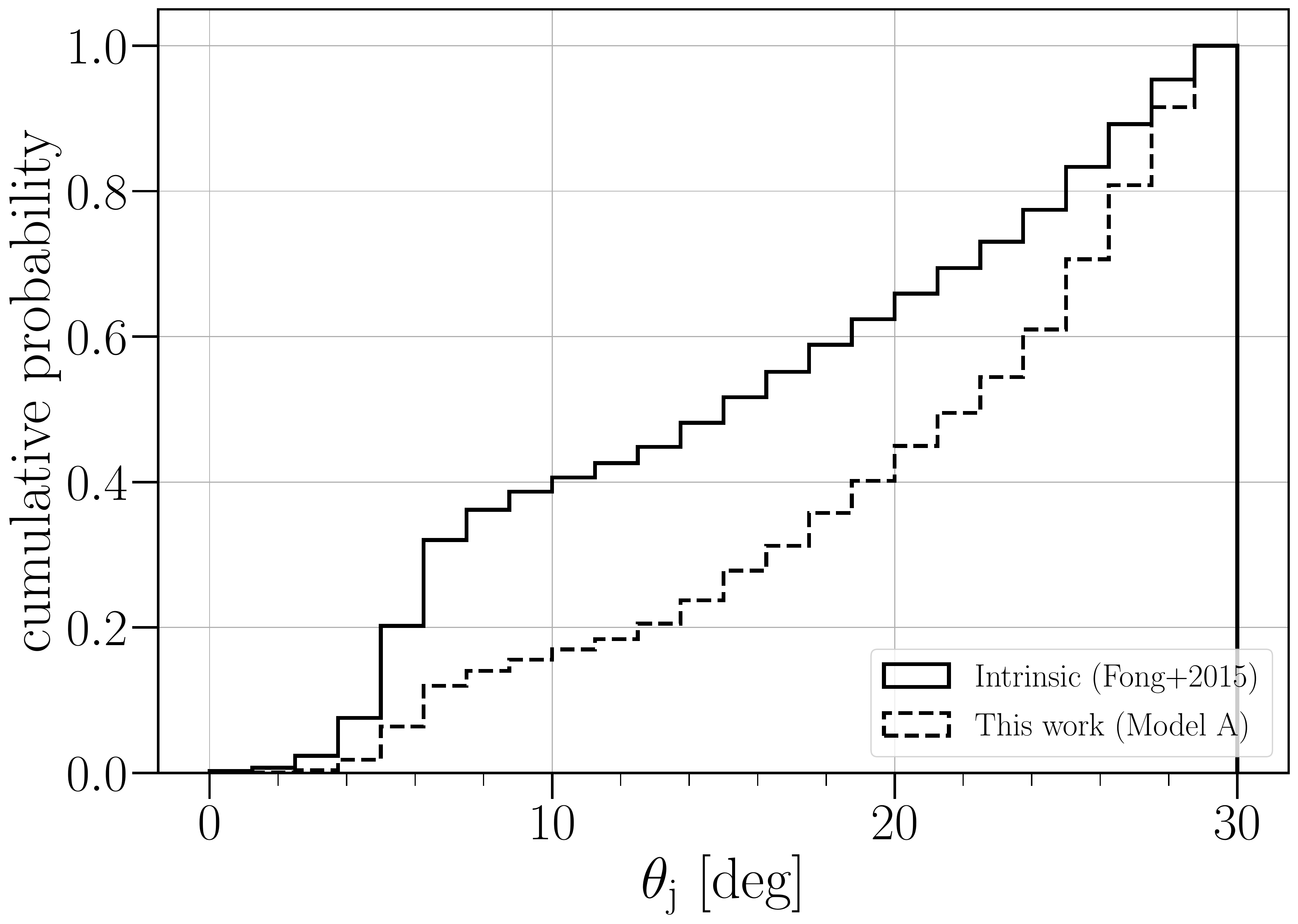}{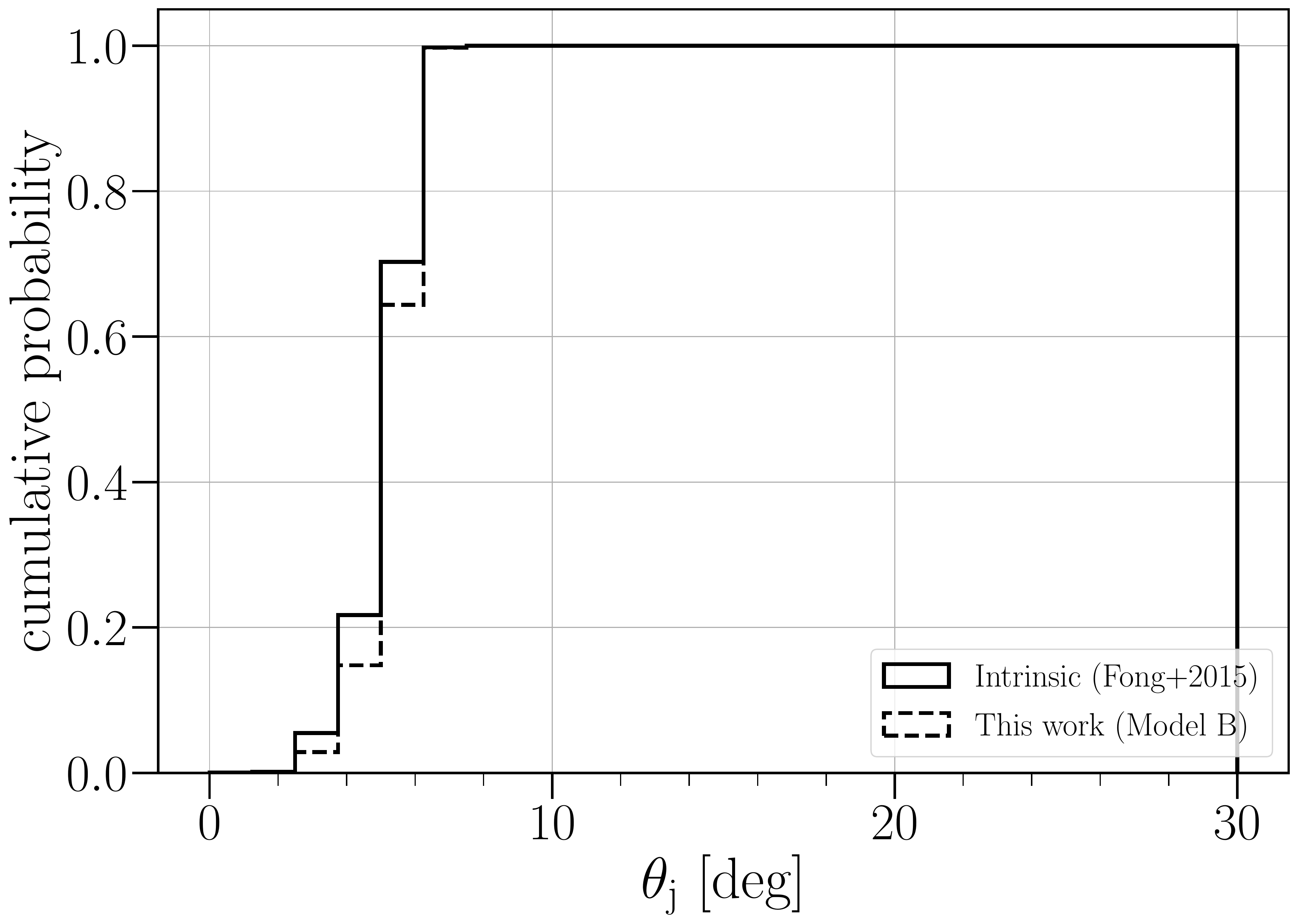}
	\caption{(Left) Cumulative density function of jet opening angle of detectable SGRBs with Model A. 
	(Right) The same except for Model B as a jet opening angle distribution. 
	In both figures, the distribution of detectable signals (dashed) is biased to larger values than the intrinsic distribution (solid) based on \cite{Fongetal2015}.} 
	\label{cum_prob_jetangle}
\end{figure}

\begin{figure}
\centering
	\includegraphics[width=\width]{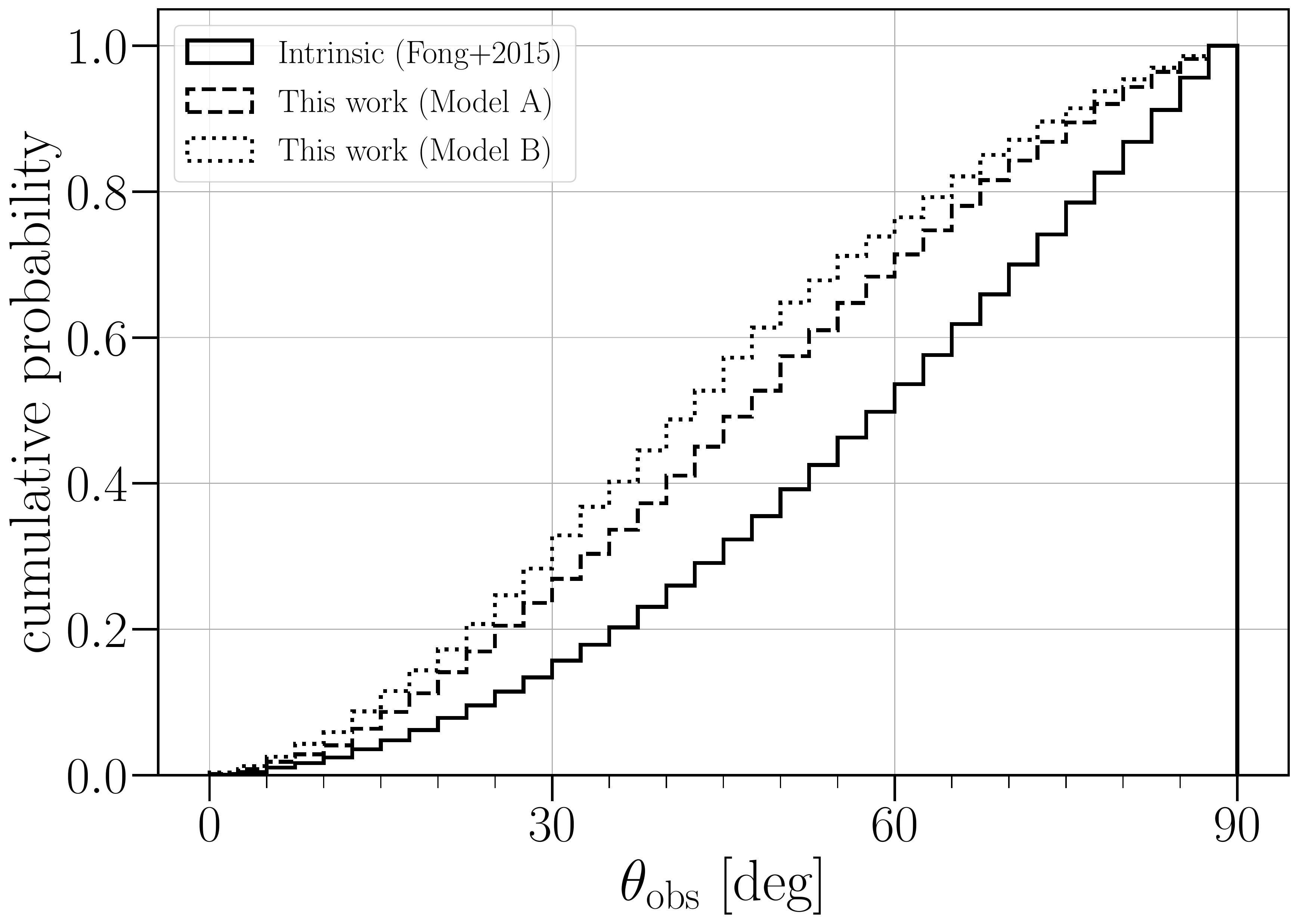}
\caption{Cumulative density function of observer angle of detectable SGRBs. The distributions of detectable signals with Model A and B (dashed and dotted histograms, respectively) are biased to smaller value than the intrinsic distribution (the solid one). Since \offaxisrationc\% and \offaxisratioc\% of the detectable signals are off-axis jet afterglows with Model A and B, and the intrinsic distributions of isotropic kinetic energy and circum medium density are based on on-axis afterglow observation, detectable signals should be more energetic than on-axis afterglows. Therefore, the distributions of isotropic kinetic energy and circum medium density of detectable signals are biased to larger values.}
\label{cum_prob_theta}
\end{figure}

Here, we summarize the parameter distribution of the detectable SGRB afterglows.
Figures \ref{cum_prob_kiso}, \ref{cum_prob_n}, and \ref{cum_prob_jetangle} are the cumulative probability densities of isotropic kinetic energy, circum medium density, and jet opening angle, respectively.  
For isotropic kinetic energy and circum medium density distributions, the dashed and dotted line histograms show our result with Model A and B respectively and the solid one is based on \cite{Fongetal2015} 
as the intrinsic distribution.
For the jet opening angle distribution, the dashed histogram is our result and the solid one corresponds to the intrinsic distribution.
For all the distributions, our results are biased to larger values compared to the intrinsic distribution, corresponding to more energetic afterglows.
The distribution of observer angle is shown in Figure \ref{cum_prob_theta}. 
Note that the viewing angle distribution is uniform on the sphere. 
We found that \offaxisrationc\% and \offaxisratioc\% of detectable afterglows have observer angle larger than the jet opening angle, $\theta_{\rm{obs}}>\theta_{\rm{j}}$, that is, off-axis ones in Model A and B respectively.
Since off-axis jet afterglows are fainter than on-axis ones, detectable off-axis afterglows should be energetic than detectable on-axis ones. Therefore, we obtained the parameter distribution biased to a larger isotropic kinetic energy, circum medium density and jet opening angle.
Off-axis SGRBs are candidates for orphan afterglows, which implies that a significant number of off-axis afterglows are detectable. Therefore, CHIME will be also effective for orphan afterglow searches.

\section{Discussion}
\label{sec:discussion}
Our result confirmed a significant number of afterglows can be detected with CHIME, $\SI{\ourrealnc}{\per\year}$ and $\SI{\ourrealc}{\per\year}$ for $3 \sigma$ confidence level with Model A and B respectively.
Here we discuss how detection rate depends on the jet opening angle. 
For top-hat jets, that is, afterglows can be observed only inside the jet cone, 
the detection probability can scale with the detectable fraction of emission solid angle, \textit{i.e.,} beaming factor $1 - \cos(\theta_\mathrm{j}) \propto \theta_\mathrm{j}^2$. 
This is true only if top-hat jets are in the relativistic regime when relativistic beaming effect is significant and thus the emission can only be observed when the line of sights is inside the jet cone of width.
When the jet eventually decelerates to non-relativistic speed, the emission becomes isotropic, and hence in late-time observations the detection probability does not scale with $\theta_\mathrm{j}$.
The start time of our light curve templates are randomly chosen from the range $[1, 365 \times 5]$\,days, 
and most of our light curve templates include only late-time emission. 
Therefore, the detectability will not change when we employ the top-hat jet model. 
However, for other jet model such as a Gaussian jet model, the detectable fraction of 
emission solid angle also depends on the detector sensitivity.
In order to verify this effect, we checked the radio flux as a function of viewing angle 
$\theta_\mathrm{obs}$ and obtain the maximum detectable viewing angle $\theta_\mathrm{max}$ 
at which the flux intersects with the sensitivity limit of CHIME. 
For computing this, we adopted the medians of other parameters such as the isotropic kinetic energy, circum medium density, and luminosity distance.
We then confirmed the ratio of $\theta_\mathrm{max}^2$ is closer to $1/3$, which is close to the detection rate ratio between Model A and B. 
That is why the detectability changes with a choice of jet opening angle distributions. 

Though \cite{TakahashiandIoka2021} showed that light curves with diverse jet structure models are consistent with the off-axis afterglow interpretation of GRB 170817A and thus the choice of different jet models should be degenerate with $E_\mathrm{k,iso}$ and $\theta_\mathrm{j}$ of a Gaussian jet, 
to what extent the detctability changes also depends on jet models. If we employ a different jet model such as a power-law jet, the dependency will change 
by a factor of $\lesssim 0.279$ with a choice of power index $b \leq 10$. 

For the direct detections, CHIME's localization is not great, and it is difficult to identify the candidates CHIME finds as SGRB afterglows and not other transients such as long GRB afterglows and active galactic nuclei (AGNs).
However, CHIME will tell us the possible regions radio transients exist. 
In follow-up observations with other radio telescopes, we can determine their positions. 
If the host galaxies are elliptical ones and the sources are located far from their center, we can exclude the possibility of long GRBs and AGNs and conclude the candidates are real signals with a high probability. 
This will hopefully supply a large enough number of samples to discuss statistically 
the parameter distribution of relativistic jet such as a jet opening angle.
Also, from radio observations, we can estimate jet energy scales without considering relativistic beaming effects. 
By determining the jet energy distribution formed by BNS mergers,
we can give some constraints on BNS mergers as the progenitors of SGRBs: whether all the BNS mergers can cause SGRBs or some of them can.
Our analysis suggests the importance of using CHIME to estimate the afterglow event rate. When the observation with CHIME takes place, three event rates will be connected for the first time:  BNS merger rate estimated by GW observation, the SGRB event rate, and radio afterglow event rate.

As \cite{TakahashiandIoka2020} derived a new method to reconstruct inversely the jet energy distribution $E(\theta)$ from an off-axis afterglow light curve, 
it is possible to estimate an energy distribution in a relativistic jet when we observe orphan afterglows. 
Even if there are no detections, as \cite{TotaniandPanaitescu2002}, \cite{Nakaretal2002} and \cite{Rossietal2008} showed that the detectability of orphan afterglows are dependent of the jet structure, 
we might be able to give a constraint on the jet structure. 
Therefore, for the actual detections, we might be able to reverse the analysis and infer the physical properties of the detected afterglows after estimating afterglow event rate.

\section{Conclusion}
\label{sec:concl}
We employed a specific detection algorithm based on likelihood ratio statistics and developed an analytic estimate of the trials factor.
We simulated the response of the detection algorithm to signals embedded in noises. Based on the result, we estimated the FAP caused by the search over sky location and the choice of light curve template. 
Taking all of that into consideration, we set a detection threshold based on target sample purity. 
We found that $3 \sigma$ confidence level is the best threshold to discuss the event rate estimations and the possibility of the direct detections. 

Considering a thermal noise and a constant background contribution, we estimated that between $\SI{\ourminnc}{\per\year}$ and $\SI{\ourmaxnc}{\per\year}$ afterglows can be detected with the median rate being $\SI{\ourrealnc}{\per\year}$ at 600 MHz with CHIME with Model A's jet opening angle distribution. For Model B, between $\SI{\ourminc}{\per\year}$ and $\SI{\ourmaxc}{\per\year}$, median $\SI{\ourrealc}{\per\year}$, are detectable. Since Model B has a smaller median value than Model A, afterglows become faint and the detectbility decreases. 
This jet opening angle dependency should change by employing a different jet model 
other than top-hat jet.
Among the detectable afterglows, \offaxisrationc\% and \offaxisratioc\% are off-axis for Model A and B respectively, which are candidates for orphan afterglows. 
When orphan afterglows are detected, physical properties of a relativistic jet can be estimated as \cite{TakahashiandIoka2020} indicated. Also, the detectability of orphan afterglows are greatly affected by the jet model and it might be possible to infer the properties of the progenitor of detectable afterglows by reversing this analysis even if no detections.
Our analysis leads to the prediction of a smaller rate of detections, but it confirms the essential conclusion of the earlier analysis \citep{Feng2014},
namely that, using CHIME to search for SGRB afterglows will be effective at constraining the astrophysical merger rate as well as searching for orphan afterglow.
Comparing BNS merger rate and SGRB event rate with the afterglow event rate should be helpful to identify the progenitor of SGRBs.
We also expect an afterglow search with CHIME can drastically increase the number of samples and discuss statistically 
the parameter distribution of a relativistic jet and we can give some constraints on BNS merger as the progenitor of SGRB.

\section*{Acknowledgement}
Our work is supported by KAKENHI 18K03692, 17H06362 and 20J12200. 

\appendix
\section{Singular Value Decomposition (SVD)}
\label{app:svd}
We have a log-likelihood ratio ranking statistic, \eqref{eq_max_lnL}, that is computed using an expression of the form
\begin{equation}
\ln \Lambda \sim (\bm{x} \cdot \bm{f})^2
\label{eq_sim_lnL}
\end{equation}
where $\bm{x}$ is a pixel time series and $\bm{f}$ is a light curve template.  We want to extremize $\ln \Lambda$ with respect to the family of functions to which $\bm{f}$ belongs, and then estimate the probability of observing such a value in a data set consisting only of noise.  We accomplish the extremiziation by computing $\ln \Lambda$ for many choices of $\bm{f}$ and picking the highest value.  As we compare the pixel time series to additional light curves, the significance of the match we will eventually identify is diminished for us having made many attempts to find it.  To estimate the false-alarm probability we need to know how many statistically independent trials we will have conducted.  Light curve templates, $\bm{f}$, that are very similar will produce similar values of $\ln \Lambda$;  knowing one of the two $\ln \Lambda$ allows one to accurately guess the other, they are not truly statistically independent trials.  What property of $\bm{f}$ makes two of them statistically independent trials for $\ln \Lambda$?  Assuming the pixel noise process to be stationary, white, and Gaussian, then when two light curves are orthogonal to each other, that is,
\begin{equation}
\bm{f}_1 \cdot \bm{f}_2 = 0,
\end{equation}
then $\bm{x} \cdot \bm{f}_{1}$ and $\bm{x} \cdot \bm{f}_{2}$ are uncorrelated.  Since they are also Gaussian random variables they are statistically independent, and therefore the $\ln \Lambda$ computed for orthogonal light curves are statistically independent.  Conversely, if $\bm{f}_1 \cdot \bm{f}_2 \neq 0$ then the $\ln \Lambda$ computed for them are not statistically independent.

The number of mutually orthogonal functions that can be found in the family of light curves tells us how many statistically independent values of $\ln \Lambda$ can be computed from that family of functions;  the $\ln \Lambda$ for any other light curve can be computed from a combination of those $\ln \Lambda$ without any need to consult the data.  We find this number by first assembling a matrix whose rows consist of light curve functions and then estimating the rank of the span of that matrix.  The matrix we constructed contained $\sim 100,000$ one-year long light curve fragments with isotropic kinetic energy between $\SI{e49}{\erg}$ and $\SI{e51}{\erg}$, circum medium density $\SI{e-5}{\centi\metre^{-3}}$, $\SI{e-3}{\centi\metre^{-3}}$, and $\SI{1}{\centi\metre^{-3}}$, observer angle \ang{0}--\ang{90} with fixed jet opening angle $\SI{0.2}{\radian}$.  Next we estimate the rank of the span of that matrix using a \ac{SVD}.  The \ac{SVD} factors a matrix into a product of two orthogonal matrices and a diagonal matrix of singular values.  The number of non-zero singular values gives the rank of the span of the original matrix.

When doing this, typically no singular values are found to be identically zero, but they will often be found to be either ``large'' or ``small'', with many orders of magnitude between the two.  An approximation of the original matrix can be obtained by replacing some number of the smallest singular values with 0s.  The vectors corresponding to the non-zero singular values that remain provide the orthonormal basis of that rank that best approximates the rows in the original matrix in the sum of square residuals sense (this is a defining property of the \ac{SVD}).  By setting all ``small'' singular values to 0, and retaining the ``large'', we obtain a sensible approximation of the light curves, and the number of singular values we retain tells us the rank of the space spanned by that approximation.  We use this for $n_{\rm{dof}}$ when estimating the false-alarm probability.

This procedure is not a rigorous derivation of the distribution of $\ln \Lambda$ extremized over the template bank, and the method by which we extract $n_{\rm{dof}}$ by counting ``large'' singular values is \textit{ad hoc}. However, the event rate is quite insensitive to the number of DOF, $n_{\rm{dof}}$:  we find that changing the estimate of this parameter by an order of magnitude in either direction changes the final estimated detectable event rate by less than 5\%. Therefore, any sensible estimate within one or two orders of magnitude of the correct trials factor is adequate.

\section{Noise Level Estimation}
\label{app:noise_est}
Here, we introduce how we derive the variance caused by antenna receiver and unresolved background sources.
First, we assume CHIME has a receiver temperature of $\SI{50}{\kelvin}$ and the sky contributes around $\SI{10}{\kelvin}$.
The contribution of noise $\sigma$ can be calculated following Equation (B5) in \cite{Feng2014},
\begin{align}
    \sigma = \left ( \frac{2 k_{\rm{B}} T}{A_{\rm{eff}} N_{\rm{ant, tot}} \epsilon_c} \right ) \frac{1}{\sqrt{N_{\rm{pol}} B t_{\rm{int}}(\delta)}},
    \label{eq:var}
\end{align}
where $T$ is a noise temperature caused by antenna receiver and the background contributor of each antenna feed, 
$A_{\rm{eff}}$ is an effective antenna area of each feed, 
$N_{\rm{ant,tot}}$ is the number of all the antenna feeds of CHIME,
$\epsilon_c$ is a correlator efficiency, 
$N_{\rm{pol}}$ is the number of polarizations, $B$ is the instantaneous bandwidth,
and $t_{\rm{int}}(\delta)$ is an integration time depending on the source declination $\delta$.
Here, we set $T=\SI{60}{\kelvin}$, $\epsilon_c=1.0$, $N_{\rm{pol}}=2$, $B=\SI{400}{\mega\hertz}$, $A_{\rm{eff}} N_{\rm{ant,tot}}=\SI{10000}{\metre\squared}$.
For $t_{\rm{int}}$, following \cite{Feng2014},
\begin{align}
	t_{\rm{int}} (\delta)
	&= \frac{1}{2 \pi} \arccos \left [ \frac{-\sin^2 \delta + \cos \Delta \cos^2 (\phi - \delta) + \sin^2 (\phi - \delta)}{\cos^2 \delta} \right ], 
	\label{tint}
\end{align}
where $\Delta = \ang{2.5;;}$ and $\phi = \ang{49;;}$.
Note that if a radio source is located at $\delta < \phi$, we will observe it once a day.
Otherwise, if it is located at $\delta > \phi$, we can observe it twice a day. For that case, the integration time $t_{\rm{int}}$ can be expressed as
\begin{align}
		t_{\rm{int}} (\delta) &= \frac{1}{2 \pi} \left [\arccos \left ( \frac{-\sin^2 \delta + \cos \Delta \cos^2 (\phi - \delta) + \sin^2 (\phi - \delta)}{\cos^2 \delta} \right ) \right . \nonumber \\ 
		&\quad + \arccos \left . \left ( \frac{-\sin^2 \delta + \cos \Delta \cos^2 (\phi + \delta) + \sin^2 (\phi + \delta)}{\cos^2 \delta} \right ) \right ].
		\label{tint2}
\end{align}
The second term of the equation includes the effect of the second transit of the source.
Averaging over the declination $\ang{0;;} \leq \delta \leq \ang{90;;}$, we obtain the variance caused by the noises thorough one day observation $\langle \sigma \rangle$ and thus the variance $\sigma^2$ can be obtained as,
\begin{equation}
	\sigma^2 = \langle \sigma \rangle^2 = (\SI{0.0232}{\milli\jansky})^2.
\end{equation}

By substituting $T=\SI{50}{\kelvin}$ for \eqref{eq:var}, we obtain an average thermal noise variance $\langle \sigma_\mathrm{th} \rangle$, 
\begin{equation}
    \langle \sigma_\mathrm{th} \rangle = \SI{0.0193}{\milli\jansky}.
\end{equation}
It is used for calculating the event rate based on Feng criterion.


\bibliography{references}{}
\bibliographystyle{aasjournal}



\end{document}